# Molecular Orientation by Intense Single Cycle THz Pulses

Sharly Fleischer, Yan Zhou, Robert W. Field and Keith A. Nelson

*Department of Chemistry, Massachusetts Institute of Technology, Cambridge, Massachusetts 02139.*

*Abstract:*

Intense single-cycle THz pulses resonantly interacting with molecular rotations are shown to induce significant field-free orientation under ambient conditions. We calculate and measure the angular distribution associated with THz-driven rotational motion and correlate the THz-induced orientation and alignment in an OCS gas sample.

In amorphous molecular systems such as gases and liquids, most spectroscopic signatures are averaged over all molecular orientations, owing to the isotropic angular distribution at the instant of measurement. Since in most light-matter processes, the interaction between the field and a molecule depends on the projection of the field polarization axis onto the transition dipole moment (whether permanent or induced), preparation of a molecular sample with well defined angular distribution prior to its further optical interrogation is beneficial in terms of efficiency, selectivity and resolution, owing to the removal or reduction of angular averaging. This understanding has motivated the development of "laser induced molecular alignment", a very active field of research in the last two decades, in which gas-phase molecules are rotationally excited by intense ultrashort laser pulses, leading to their field-free, periodically recurring, highly squeezed, angular distributions (alignment) and serving as a preparation step for further interactions, such as strong field ionization [1] and dissociation, scattering off surfaces [2,3,4], deflection by inhomogeneous fields [5], high harmonic generation (HHG) [6,7,8,9], measurement of molecular frame photoelectron angular distributions [10,11], ultrafast X-ray diffraction (UXD) [12,13], and many others. However, the nonresonant nature of this field-molecule interaction preserves the up-down symmetry of the medium during and following the interaction, so no net orientation of asymmetric molecules is induced in the medium. Symmetry breaking and net molecular orientation has been achieved through the use of DC fields that interact with the molecular permanent electric dipole [14,15]. However, the presence of a DC field may affect the results of measurements performed on oriented samples and thus it is of special importance to achieve molecular orientation under field-free conditions as was demonstrated recently through the use of moderate quasi-DC fields followed by pulsed optical excitation of low density, jet-cooled molecular samples [16,17,18]. Another approach used a two-color optical scheme in which the fundamental laser frequency was mixed with its second harmonic [19]. In these cases, however, the molecules are exposed to relatively intense optical irradiation ($\sim 10^{13}$ W/cm$^2$) prior to their interrogation, potentially producing unwanted molecular excitation (which would be a severe disadvantage for biological samples due to their relatively low damage thresholds).

In this work we demonstrate significant field-free orientation of polar gas phase molecules by intense single-cycle THz fields at ambient conditions. We measure both the time-dependent orientation and alignment by probing the free-induction-decay (FID) and the transient birefringence induced by THz excitation. The measurements exploit recently achieved generation of THz pulses with microjoule energies [20,21], which has enabled nonlinear THz spectroscopy in solid [22,23,24], liquid [25], and now gas phase samples.

Polar molecules can interact with EM fields that are resonant with their rotational transition frequencies via their permanent dipole moments. Classically, the linearly polarized field, $\vec{E}(t)$ acts to orient the initially isotropic molecular dipole vectors, $\vec{\mu}$, with the field direction through the interaction potential, $V(\theta,t) = -\vec{\mu} \cdot \vec{E}(t) = -\mu E(t)\cos(\theta)$, where $\theta$ is the angle between the polarization and dipole vectors. This exerts a torque on the molecules and initiates coherent rotational motion. Quantum mechanically, a resonant field couples adjacent opposite parity rotational states with $\Delta J = \pm 1$, $\Delta m = 0$, transferring population and inducing coherences that give rise to the emission of FID signals [26,27]. Excitation with a weak single-cycle THz pulse, the broad bandwidth of which includes many rotational transition frequencies, $2nBc$, where $B$ is the molecular rotational constant, $c$ is the speed of light in vacuum, and $\{n\}$ are positive integers (i.e. a THz field the single period of which is short compared to the associated rotational periods) produced coherences at all of the resonant transition frequencies [28,29]. Interferences among these single-quantum coherences suppressed far-field emission, except when they all radiated in phase at equally spaced bursts termed "commensurate echoes" following the THz excitation pulse with a "revival" period of $T_{rev} = 1/2Bc$. The far-field THz emission was detected at the times when the molecules undergoing coherent rotation came into transient net orientation of their dipoles, described by the ensemble average $<<\cos\theta>>$. This periodic behavior is closely reminiscent of the "quantum rotational revivals" [30,31,32,33] that follow nonresonant femtosecond optical excitation of two-quantum rotational coherences through impulsive stimulated rotational Raman scattering, also known as "laser induced molecular alignment". A short nonresonant pulse acts on the molecular polarizability anisotropy, $\Delta\alpha = \alpha_{\parallel} - \alpha_{\perp}$ (where, for linear molecules, as in our case, $\alpha_{\parallel}$ and $\alpha_{\perp}$ are the polarizability components respectively parallel and perpendicular to the molecular axis) through the interaction potential, $V(\theta,t) = -\Delta\alpha E^2(t)\cos^2(\theta)$, that couples rotational levels with $\Delta J = \pm 2$, $\Delta m = 0$. These two-quantum coherences ($\rho_{j,j\pm 2}$) do not radiate, but they can be detected when they are in phase (twice as often as the revival period defined above, i.e. at time intervals of $T_{rev}/2$, corresponding to transient net alignment and anti-alignment, described by the ensemble average, $<<\cos^2\theta>>$) as induced birefringence (essentially coherent depolarized Raman scattering between rotational levels with $\Delta J = \pm 2$) of a variably delayed optical probe pulse.

In the present work we excite not just one-quantum rotational coherences but also multiple-quantum coherences through repeated interactions between a strong THz field and the sample, and we detect, under field-free conditions, both the FID commensurate echoes observed earlier after weak THz excitation [28,29,37] and the two-quantum rotational revivals previously observed only after intense optical excitation [32,33].

We used 6-mJ optical pulses to generate ~ 4-µJ single-cycle THz pulses as described previously [20, 21]., About 2 µJ of the THz pulse energy reached the sample, focused to a spot diameter of ~1.5 mm. The path length in the gas was 10 cm, and the sample was at room temperature for all measurements (See Fig. 1). The transmitted THz field and the induced THz FID bursts or "commensurate echoes" that followed were measured by electro-optic sampling, i.e. through the birefringence that the THz field induced in a ZnTe nonlinear crystal, as monitored by a variably delayed optical readout pulse. The measured THz echoes revealed the induced molecular orientation, $<<\cos\theta>>(t)$. Alternatively, the induced molecular alignment, $<<\cos^2\theta>>(t)$, was determined by measurement of the corresponding induced birefringence in the gas, monitored by a variably delayed optical probe pulse.

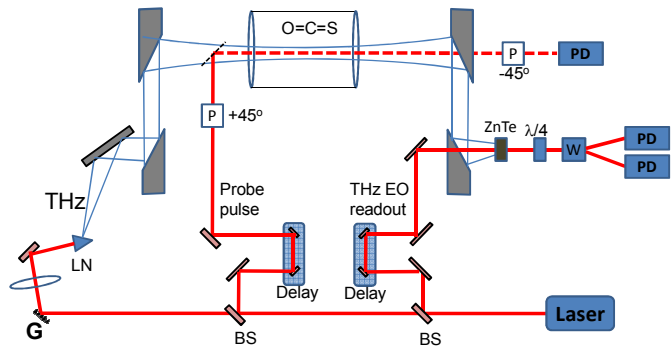

Figure 1. Schematic representation of the two setups used in this work. An optical pulse was used to generate the single-cycle THz pulse that was passed through the gas cell to induce coherent rotational motion. Either the THz electro-optic (EO) readout pulse was generated for measurement of the periodically emitted THz fields due to the induced molecular orientation, or the optical probe pulse was generated for measurement of the THz-induced birefringence due to molecular alignment. **G**, grating; **LN**, lithium niobate THz generation crystal; **BS**, beamsplitter; **P**, polarizer; **W**, Wollaston prism; **PD**, photodetector.

Figure 2a depicts a THz pulse and FID emission induced by it due to molecular orientation of OCS at 250 torr.

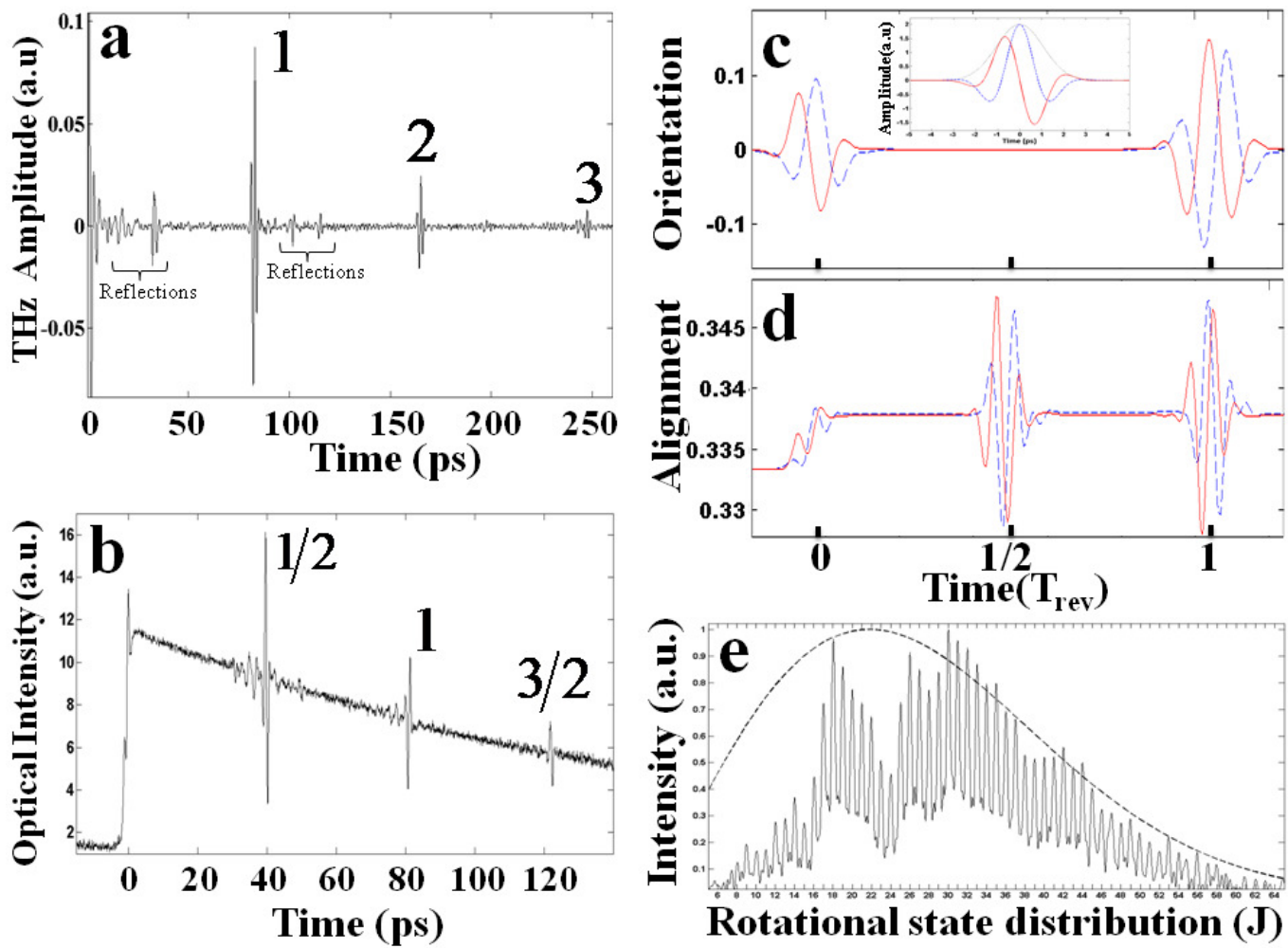

**Figure 2:** (a) Electro-optic sampling of the FID emission from the OCS gas cell following the THz pulse. (b) Induced birefringence of the OCS gas cell. (c) Numerical simulation results for orientation $<<\cos\theta>>(t)$ induced by a single-cycle THz pulse. Two simulations are shown for THz pulses with the same pulse envelope (inset, black dotted curve) and with fields phase-shifted by 90° (inset, solid red and dashed blue curves). (d) Numerical simulation results for alignment $<<\cos^2\theta>>(t)$ induced by a single-cycle THz pulse. Simulations are shown for the same phase-shifted THz waveforms as in (c). (e) Normalized rotational level ($J$ state) contribution to the coherent signal shown in (b). The calculated thermal rotational distribution at 300 K is depicted by the dashed curve.

Following the intense THz pulse at $t = 0$, the data show signals at ~ 22 ps due to a reflection in the ZnTe EO sampling crystal and at ~ 30 ps due to a reflection from the plastic (TOPAS) window at the front of the gas cell. The first three revivals of the OCS are apparent at $T_{rev} = 82\,ps$ intervals (marked as 1,2,3 in Fig.2a). The peak amplitude of the OCS FID at the first revival (0.1 a.u.) is ~15% of the peak amplitude of the driving field (0.65 a.u., not shown), suggesting a significant orientation response induced by the THz field.

Figure 2b shows the time-dependent birefringence due to THz-induced alignment in the OCS gas at 350 torr. The THz pulse was applied at time $t = 0$. The two peaks, located at ~ 1.3 ps and 0 ps, originate from an instantaneous electronic contribution to the birefringence. The higher peak at t=1.3 ps results from the contribution of the alignment induced by the first half-cycle of the pulse as well as the instantaneous electronic response to the second half-cycle to the observed signal. Strong revivals are observed at 41 ps, 82 ps and 123 ps (1/2, 1, 3/2 $T_{rev}$, respectively) with inversion between successive signal profiles, a familiar feature of two-quantum rotational coherences [34,35]. Also apparent is an induced background signal level that reaches its maximum value at around the end of the THz pump pulse, $t \sim 2$ ps, and decays slowly thereafter. This background change is a manifestation of a non-thermal rotational state distribution that is the result of population transfer to higher rotational states, giving rise to anisotropic angular distribution. Thus the total extent of alignment $<<\cos^2\theta>>(t)$ can be expressed as a sum of two parts, which are due to coherences $<<\cos^2\theta>>_c(t)$ and populations $<<\cos^2\theta>>_p(t)$ [36].

Figures 2c and 2d show the results of numerical simulations of time-dependent molecular orientation, $<<\cos\theta>>(t)$, and alignment, $<<\cos^2\theta>>(t)$, induced by a THz pulse with field profile $E(t)$ consisting of a single sinusoidal or cosinusoidal cycle. In our calculations, we used the density matrix formalism and numerically solved the Liouville-Von Neumann equation, $\frac{\partial \rho}{\partial t} = -\frac{i}{\hbar}[H, \rho]$, where $H = \frac{\hat{L}^2}{2I} + V(\theta, t)$ is the Hamiltonian, $\hat{L}$ is the angular momentum operator, $I$ is the moment of inertia, and $V(\theta, t) = -\mu E(t) \cos(\theta)$ is the interaction potential, as discussed above. A reduced sample temperature, $T = 50$ K, was chosen to reduce the number of thermally populated rotational states that needed to be considered in the calculation. The results were qualitatively identical to those at higher temperatures, but the calculation time was reduced significantly.

The molecular orientation depicted in Fig. 2c shows the expected recurrences separated by the quantum revival time ($T_{rev}$). The transient orientation (and corresponding emitted THz field) profile at the first recurrence is 90° phase shifted with respect to the incident THz field profile. All of the subsequent recurrences at integer $T_{rev}$ (not shown) are similar to the first. A 90° phase shift in the incident THz field (the red and blue curves are 90° shifted with respect to one another) results in a corresponding phase shift in the induced orientation and the THz field emitted by it. The calculated time-dependent molecular alignment, shown in Fig. 2d, reveals recurrences separated by $T_{rev}/2$, indicating, in view of Fig. 2c, that at half-revival times the sample shows orientationless alignment (thus retaining up-down symmetry), while at integer revival times the sample shows both net alignment and net orientation. Successive alignment profiles are inverted, and the 90° phase shift in the driving field (red vs. blue) results in a 180° phase shift in the alignment profiles. The simulation also shows the steady-state background signal level due to THz-induced changes in the $J$-state population distribution, however it does not include relaxation processes [36], which are outside the scope of the present paper.

A Fourier transform of the time-domain alignment signal in Fig. 2b (after removal of the exponentially decaying background) yielded a succession of features at evenly spaced frequencies, which, when rescaled to the two-quantum rotational coherences ( $f_{J,J+2} = (4J+6)Bc$ ) of the OCS molecule $B = 0.203\ cm^{-1}$ [37], revealed the contributions of more than 50 rotational states to the observed signal, as shown in Fig. 2e. Comparison to the initial thermal population shows that the THz pulse is sufficiently broad spectrally to induce coherences involving essentially all of the thermally populated rotational states, although, at the lower frequencies, the THz spectral content and the corresponding coherence amplitudes are relatively weak.

Finally we estimate the magnitudes of the orientation and alignment that produced our detected signals. As has been discussed previously [36], the time-dependences of the two contributions to alignment, $<<\cos^2\theta>>_p(t)$ and $<<\cos^2\theta>>_c(t)$, can be approximated as single-exponential decays. We find the corresponding relaxation times to be $T_1 = 172$ ps and $T_2 = 57$ ps, and extrapolation to time $t = 0$ yields the amplitudes, $<\cos^2\theta>_p(0) = 11.6$ a.u. and $<\cos^2\theta>_c(0) = 25.7$ a.u., respectively for the data in Fig. 2b, with the $<\cos^2\theta>_c(0)/<\cos^2\theta>_p(0) = 2.2$ ratio in excellent agreement with our numerical simulations.

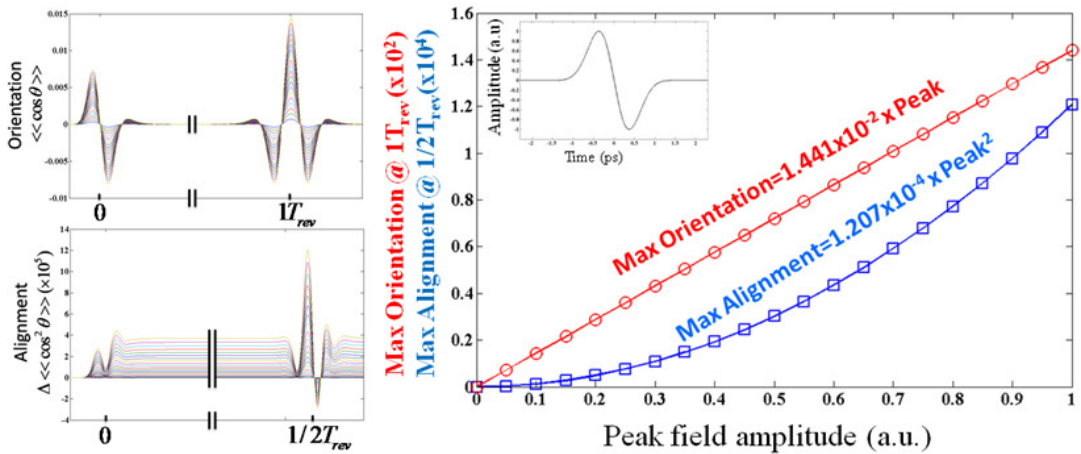

**Figure 3: (a) Time-dependent orientation and alignment induced in OCS at 300K, calculated for a series of amplitudes of the driving field profile depicted in the inset of panel b. (b) By fitting the peak orientation and peak alignment we get their peak amplitude dependences marked in the figure.**

The relation, $\frac{\Delta I}{I} = \sin(\frac{\Delta n \omega L}{c})$, correlates the observed signal level, $\frac{\Delta I}{I} = 9 \cdot 10^{-3}$, and the birefringence $\Delta n$ induced in our experiment, yielding $\Delta n = 1.78 \cdot 10^{-8} = \frac{3N\Delta\alpha}{4n\varepsilon_0}(\Delta <\cos^2\theta>)$ [38], from which the maximal change in alignment factor induced in our experiment is calculated to be $\Delta <<\cos^2\theta>> = 3.8 \cdot 10^{-5}$. In order to extract the degree of orientation induced, we used the degree of alignment determined experimentally, simulated the induced alignments and orientations at various field strengths, and correlated between the two observables as presented in figure 3. The results yielded a maximal experimental orientation factor of 0.0086 (~1%). We note that the values calculated above are averaged over the entire 10 cm gas cell length. By considering the spatial profile of the THz pulse throughout its propagation in the cell, we find the maximal orientation that occurs near the focal plane to be ~5 times higher (5%). Averaged over the cell length, the THz field amplitude was ~ 22 kV/cm, which is 5-10 times smaller than our routinely generated peak fields [20] and far weaker than the ~ 1 MV/cm fields we can now generate and has been reported recently [39]. We further note that the use of a glass front cell window to enable copropagation of the THz pump and optical probe pulses results in loss of ~30% of the THz field. Therefore far higher degrees of alignment and orientation are expected with more optimal excitation conditions. Jet cooled molecular samples will allow still stronger responses.

Intense single-cycle THz pulses have been used to induce a significant degree of orientation in gas phase polar molecules. With realistic THz excitation conditions, optical measurements, such as HHG, X-ray diffraction, and ion-fragmentation imaging, can be performed on oriented samples and under field-free conditions. The combination of stronger THz fields (acting on the permanent dipole) and optical fields (acting on the anisotropic polarizabillity tensor) will offer two independent handles for controlling the 3D angular distribution of more complex (nonlinear polyatomic) molecules.